\documentclass[]{scrartcl}
\usepackage{graphicx}
\usepackage{natbib}
\usepackage{subcaption}
\usepackage{booktabs}
\usepackage{amsmath}
\usepackage{hyperref}

\usepackage{dcolumn}
\usepackage{booktabs}
\usepackage{tikz}
\usetikzlibrary{positioning,shapes,arrows, fit}
\usepackage{pgfplots}
\pgfplotsset{compat=1.6}
\newcolumntype{M}[1]{D{.}{.}{1.#1}}
\usepackage{amsmath}

\usepackage{graphicx}

\usepackage{geometry}
\newgeometry{vmargin={25mm}, hmargin={18mm,18mm}}

\usepackage{amssymb}
\usepackage{pifont}

\usepackage{authblk}

\title{\vspace{-1.5cm} \noindent\Large\textbf{Dependent Infrastructure Service Disruption Mapping (DISruptionMap): \\ A Method to Assess Cascading Service Disruptions in Disaster Scenarios}}

\date{\vspace{-7ex}}

\small
\author[1, *]{\small Moritz Schneider}
\author[1]{\small Lukas Halekotte}
\author[1]{\small Andrea Mentges}
\author[2]{\small Frank Fiedrich}
\affil[1]{\small German Aerospace Center (DLR), Institute for the Protection of Terrestrial Infrastructures, Germany}
\affil[2]{Chair for Public Safety and Emergency Management, University of Wuppertal, Germany} 
\affil[*]{Corresponding author, e-mail: moritz.schneider@dlr.de} 

\begin{document}
\maketitle

\par\noindent\rule{\textwidth}{0.3pt}
\begin{abstract}
Critical infrastructures provide essential services for our modern society. 
Large-scale natural hazards, such as floods or storms, can disrupt multiple critical infrastructures at once. 
In addition, a localized failure of one service can trigger a cascade of failures of other dependent services.
This makes it challenging to anticipate and prepare adequately for direct and indirect consequences of such events. 
Existing methods that are spatially explicit and consider service dependencies currently lack practicality, as they require large amounts of data. 
To address this gap, we propose a novel method called \textit{DISruptionMap} which analyzes complex disruptions to critical infrastructure services. 
The proposed method combines i) spatial service models to assess direct service disruptions with ii) a service dependency model to assess indirect (cascading) service disruptions. 
A fault tree-based approach is implemented, resulting in a significant decrease in the information required to set up the service dependency model.
We demonstrate the effectiveness of our method in a case study examining the impact of an extreme flood on health, transport, and power services in Cologne, Germany. 
\end{abstract}
\par\noindent\rule{\textwidth}{0.3pt}


\section{Introduction}
Critical infrastructures (CIs) form the backbone of modern societies \citep{Nick2023}, providing them with essential services, including mobility, electricity, and healthcare. These services are increasingly exposed to a growing number of natural hazards \citep{Ward2020, Merz2020}, such as floods, storms, or earthquakes. Such large-scale disruptive events often damage multiple CIs simultaneously, e.g. a flood might cause disruptions in the road network, the energy distribution system, and the healthcare system. As CIs often span over or supply large geographical areas \citep{Arvidsson2023}, CI failures can lead to service impairments far away from the immediate location of the actual disruptive event. Additionally, modern CI systems can be subject to complex indirect hazard impacts: CIs are highly interdependent across different sectors \citep{rinaldi2001, Nick2023}, e.g. a power plant depends on a steady supply of cooling water and in turn provides electricity to numerous other CI systems. These dependencies can be of various types (e.g. physical, cyber, informational, political \citep{Rehak2018} or service centered \citep{Stergiopoulos2016, HallMay2010}) and can cause indirect disruptions due to the failure of one infrastructure inducing the failure of other subsequent infrastructures -- often referred to as cascading effects \citep{Arvidsson2023}. Due to these complex dependencies, the extent and consequences of CI disruptions are afflicted with significant uncertainties, which makes their prediction highly challenging.

Several approaches have been put forward to describe the direct impact of natural hazards on CIs. Multiple works focus on spatial models of direct CI disruptions caused by e.g. storms \citep{Haraguchi2016}, floods \citep{Dawod2012}, earthquakes \citep{Tamaro2018}, or explosions \citep{Fekete2023}. These works use GIS-based overlay analyses of hazard exposure areas and CI locations to estimate the impact in affected areas. Some of these works also consider uncertainties in CI failures depending on the intensity of CI component exposure, e.g. modeled by fragility curves (see \cite{SerranoFontova2023} for an overview). Other works focus on the (quantitative) modeling of CI dependency patterns and associated cascading effects \citep{Gong2023,Fekete2019}, with some also accounting for uncertainties in their analyses \citep{Grafenauer2018,Rehak2018,Stergiopoulos2016}. These approaches differ with regard to the modeling technique, e.g. there are network models, input-output-models, multi-agent systems, or Bayesian networks \citep{Arvidsson2023}. 

Few studies combine spatial CI exposure assessments with CI dependency models \citep{Schotten2023,Arvidsson2023,Gordan2024,Muehlhofer2023}.
Such methods to assess (cascading) CI disruptions usually rely on a potentially high volume of different types of data: geo-referenced data of CI locations and hazard distribution for a specific scenario \citep{Fekete2020}, data for the assessment of direct CI disruptions \citep{SerranoFontova2023}, and data for indirect CI disruptions \citep{Rehak2018}. 
As suitable historical data on severe disruptive events that caused serious cascading failures is scarce, researchers often need to rely on expert knowledge. 
The need to effectively integrate expert knowledge results in a characteristic area of tension:
on the one hand, a certain level of detail is required for an accurate description of the CI dependency network \citep{Schotten2023}.
On the other hand, the burden on individual experts to deliver large amounts of data should be limited, to ensure practicality of the method -- an aspect that is often lacking in this context \citep{Arvidsson2023}.
The level of detail thus needs to be balanced with the simplicity of the modeling approach. 

To address this, we developed a method called Dependent Infrastructure Service Disruption Mapping (\textit{DISruptionMap}). 
\textit{DISruptionMap} enables a spatial assessment of direct as well as indirect CI service disruptions and allows the effective integration of  expert knowledge while requiring a minimum amount of information.
It is composed of GIS-based spatial models (to assess the direct CI disruptions) and a BN-based service dependency model (to assess indirect CI disruptions). 
BNs \citep{pearl1985bayesian} are powerful tools to model CI dependencies and associated uncertainties \citep{DiGiorgio2012,HOSSAIN2020}. 
They can be constructed solely based on expert knowledge, but also using a combination of expert knowledge and other data sources.  
Here, we construct a BN by describing CI dependencies as services, such that the dependencies can be consistently and uniformly defined across different types and sectors of CIs. 
The resulting information on (cascading) service disruptions throughout the affected area can be used by practitioners for planning of disaster response measures or training exercises. 

The remainder of this paper is organized as follows: first, we provide an overview of the \textit{DISruptionMap} method, highlighting the main steps for its application (for a more detailed description, see "Methods"); second, we demonstrate the method through a case study of a 500-year flood scenario in the city of Cologne (Germany), examining disruptions to multiple CI services with a focus on hospital emergency care services; finally, we discuss limitations of our proposed method and the opportunities it offers for future research and for the application in disaster management.

\section{DISruptionMap: Dependent Infrastructure Service Disruption Mapping}
The \textit{DISruptionMap} method requires several steps of preparation: 
i) determine the specific hazard causing the disaster scenario (e.g. a 500-year river flood event), along with an impact measure (IM) to quantify its severity (e.g. water depth for a flood); 
ii) find Geospatial data which shows the expected IM intensity levels within the study area, e.g. a map showing the flood depth;  
iii) identify a target critical infrastructure (TCI) that describes the service of central interest and serves as a starting point for a top down approach to reflect (T)CI dependencies in the study area; and
iv) collect spatial data on TCI locations and CI components, with the type of CIs first  being identified during the development of the service dependency model.
Given these four bits of information (the specific disaster scenario and its IM, geospatial IM intensity data in the study area, the TCI, and spatial data on the (T)CI), the \textit{DISruptionMap} method can be applied.

\begin{figure}[!ht]
	\centering \includegraphics[width=10cm]{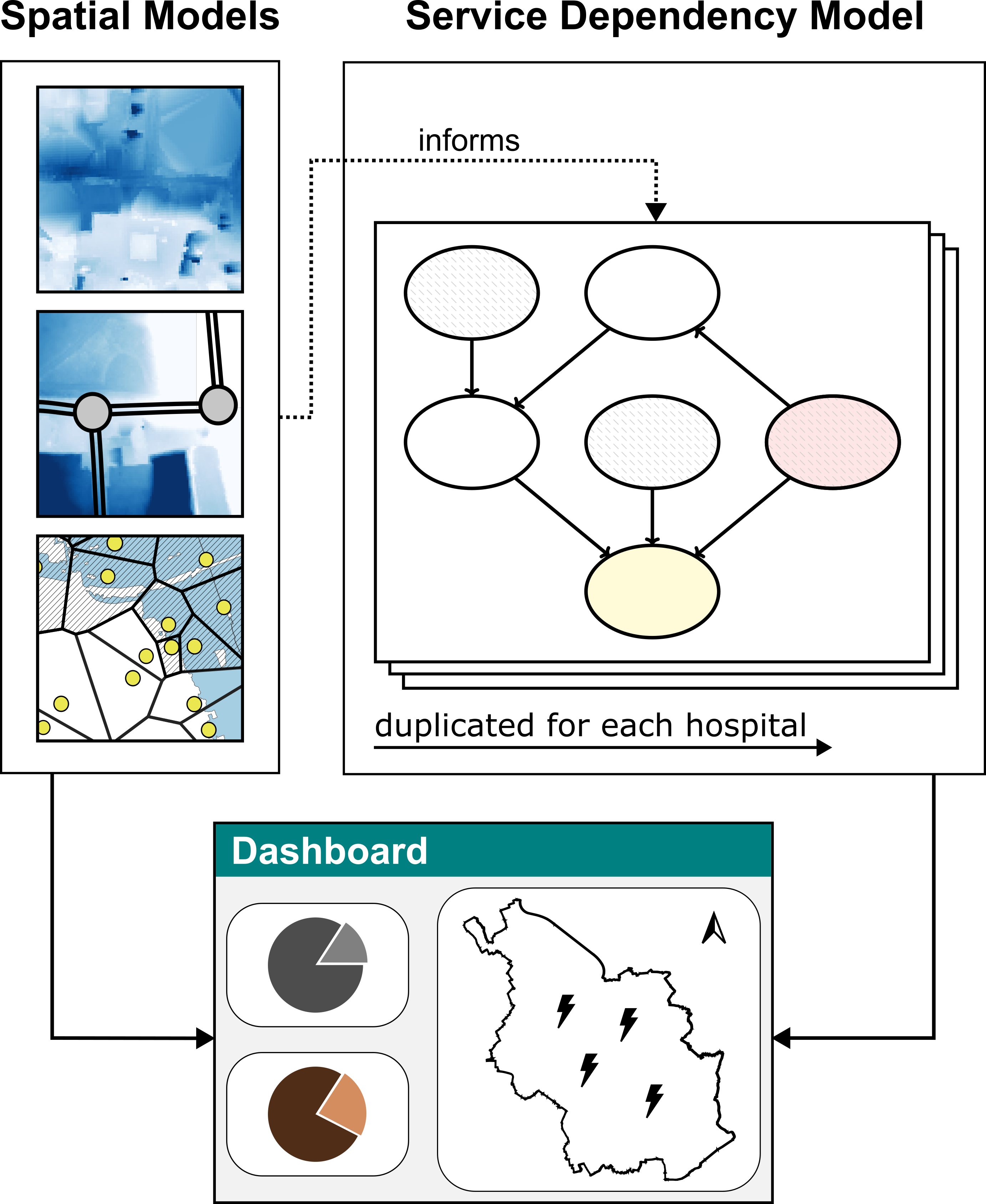}
	\caption{Summary of the proposed method, using the case study as an example. The proposed method combines GIS-based spatial service models with a BN-based service dependency model (i.e. GIS informs BN). The results are displayed in an interactive dashboard to help gain a quick overview of the high volume of (spatial) information. }
	\label{method_overview}
\end{figure}

The key steps to apply the proposed method are (see "Methods" for details, and Fig. \ref{method_overview} for an overview in the context of the case study): (i) Building \textit{Spatial Service Models}, i.e. develop spatially explicit models to assess direct CI service disruptions based on the hazard-specific IM intensity map, and (ii) Developing the \textit{Service Dependency Model}, i.e. a Bayesian network-based CI service dependency model to assess indirect disruptions. This service dependency model is replicated for each TCI in the study area to individually assess the (combined) disruptions on the TCI service, e.g. the TCI is a hospital and thus the service dependency model is duplicated and assigned to each hospital in the study area. In combination, the two models (spatial and dependency models) allow to assess both direct and indirect CI service disruptions within the study area. Finally, the results are displayed in an interactive dashboard, providing a both concise and comprehensive summary to the large volume of (spatial) information on the service disruptions for all entities in the study area. 

\section{Case Study of a Flood Scenario in the City of Cologne} \label{case_study}
\begin{figure}[!ht]
	\centering \frame{\includegraphics[width=10cm]{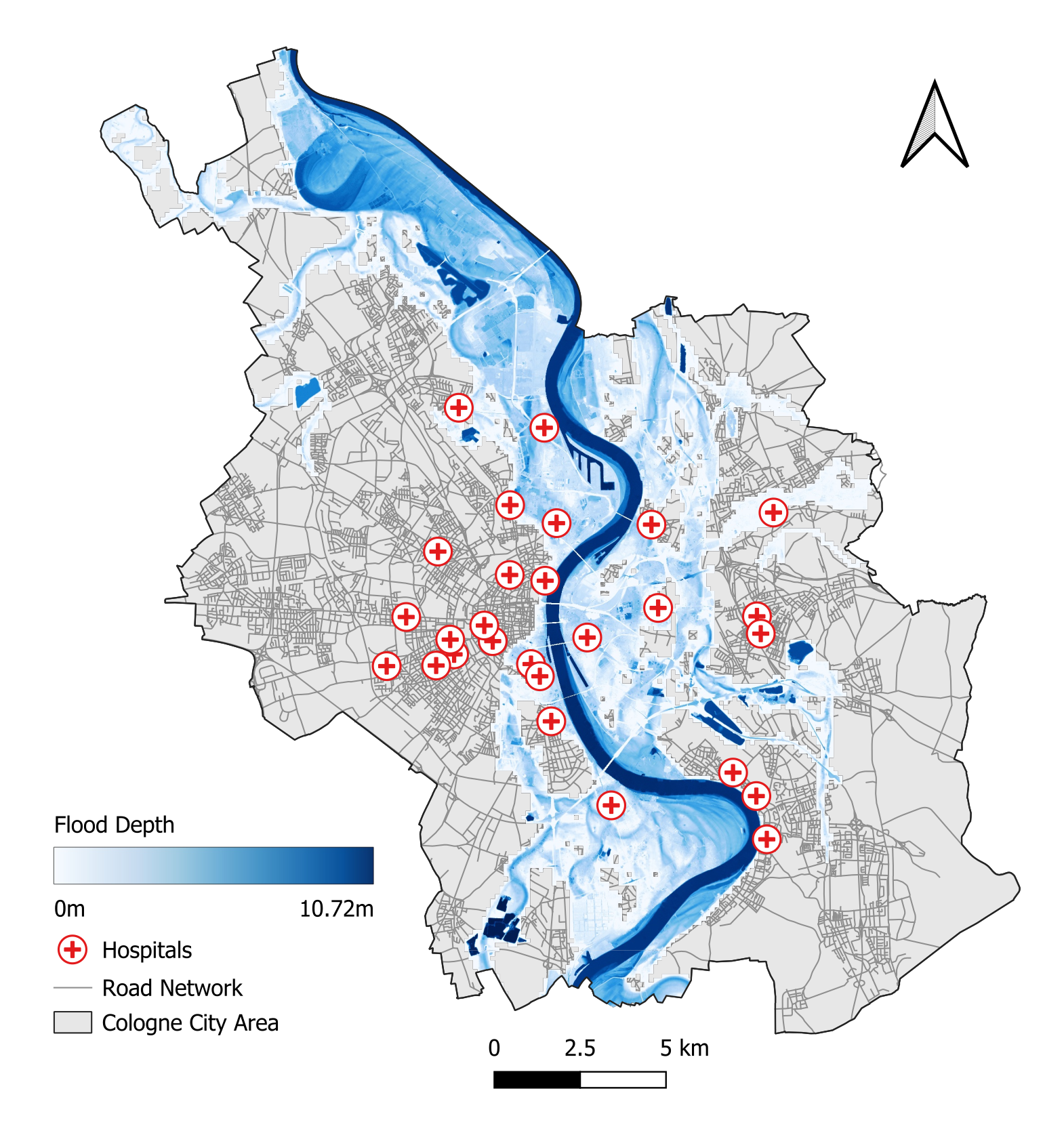}}
	\caption{Map of case study region. The city of Cologne is displayed along with the locations of hospitals, the road network, and the flood depth in a scenario of a 500-year flooding (HQ500).}
	\label{scenario}
\end{figure}

We illustrate our proposed method in a case study conducted in Cologne (Germany), focusing on emergency care service availability by hospitals during an extreme river flood scenario (Fig. \ref{scenario}). Cologne, one of Germany's largest cities with over one million inhabitants, is vulnerable to river floods: the Rhine river runs right through the city center. The city has experienced several devastating floods, e.g. in 1993 and 1995, with significant impacts on citizens and economic damages \citep{Fink1996, nhess-10-1697-2010}. Moreover, a flood event that occurred in the Ahr Valley (Germany) in 2021 further underscored the importance of effective flood risk management in this region \citep{Bier2023}. Cologne's dense population, immediate proximity of the city center to the Rhine river, and high concentration of critical infrastructures make it a relevant and complex study area for testing the proposed method. 

The GIS data used in this case study stems from a hydrological simulation of an extreme flood scenario (also called 500-year flood, see Fig. \ref{scenario}) that is provided by the state of North Rhine-Westphalia. It includes failed or overtopped dikes, mobile flood defenses, and groundwater intrusion into old river arms \citep{Fekete2020}. The analysis of such extreme flood scenarios is mandatory under the European flood directive and is available for all major German rivers \citep{Fekete2020}. Furthermore, we use CI location data from Open Street Map (OSM) for the spatial service models. Note that the data retrieved from OSM is not checked for accuracy and completeness and can thus not be transferred directly to the real-world infrastructures in the study area. The case study is intended as a proof of concept of the proposed method using example data retrieved from OSM.

In the following, first, we describe the development of the service dependency model (implemented in Python using the pgmpy library \citep{ankan2015pgmpy}); second, we detail the development of the spatial service models (using the geopandas library \citep{kelsey_jordahl_2020_3946761}); and finally, we present the results and visualize them in a dashboard environment (using the HoloViz library).

\subsection{Service Dependency Model}
\begin{figure}[!ht]
	\centering \includegraphics[width=10cm]{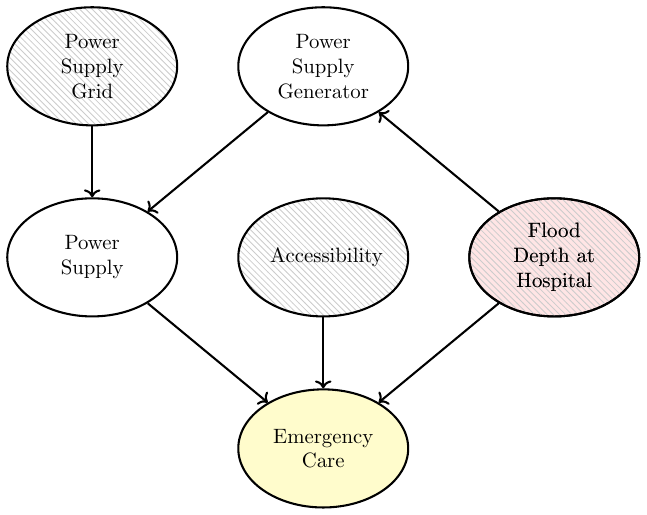}
	\caption{Bayesian network-based service dependency model of the case study. The model includes six nodes including the target node \textit{Emergency Care} highlighted with yellow fill. Leaf nodes that are informed by the spatial models are highlighted with dashed fill. The node directly informed by the hazard map is highlighted with red fill.}
	\label{bn_study}
\end{figure}

In this case study, we examine hospitals as target critical infrastructures (TCIs) and their dependencies on other services. In specific, we focus on the availability of \textit{emergency care} addressing services provided to patients who require immediate attention due to a sudden or unexpected condition. Thus, the target node of the BN-based service dependency model is \textit{Emergency Care}. To quantify the direct impact of the flood scenario on the emergency care services of an individual hospital, the impact measure node \textit{Flood Depth at Hospital} is introduced as a parent node of the target node (Fig. \ref{bn_study}). As highlighted in previous research \citep{Gordan2024}, hospitals crucially rely on power supply (for emergency care services) and road network accessibility (for emergency vehicles and goods supply). Therefore, the services \textit{Accessibility} and \textit{Power Supply} are introduced as parent nodes of node \textit{Emergency Care}. In Germany, two redundant power supply sources are mandatory for hospitals: supply by the power grid and a backup generator which is located within the hospital. Therefore, we introduce two additional parent nodes to the node \textit{Power Supply}: \textit{Power Supply Grid} and \textit{Power Supply Generator}. The power generator is located within the hospital and thus it can also be considered as a child node of node \textit{Flood Depth at Hospital}. 
Ultimately, the full BN (Fig. \ref{bn_study}) describing all dependencies of the target node \textit{Emergency Care} consists of three dependent nodes (\textit{Emergency Care, Power Supply,} and \textit{Power Supply Generator}) that each require one Conditional Probability Table (CPT). For the node \textit{Flood Depth at Hospital}, we distinguish between four categories of flood depths. All other nodes are treated as binary, i.e. they show two states (Table \ref{states}).

Two categories of nodes within the dependency model are informed by spatial information: for the direct disruption, \textit{Flood Depth at Hospital}, which measures the flood depth at the immediate location of the hospital itself, and, for the indirect disruptions, \textit{Power Supply Grid} as well as \textit{Accessibility}, which analyze service failures caused by the flooding of transformer stations or roads at a distance from the hospital. For instance, a flooded road segment 1$\,$km away from the hospital could render it inaccessible. 

\begin{table}[htbp] 
  \centering  
  \resizebox{\textwidth}{!}{%
    \begin{tabular}{|c|c|c|c|c|c|} 
    \hline
    \textbf{Flood Depth (D)} & \textbf{Accessibility} & \textbf{Emergency Care} & \textbf{Power Supply} & \textbf{Power Supply} & \textbf{Power Supply} \\
    &&&&\textbf{Generator}&\textbf{Grid} \\
    \hline
    None (0m) & Accessible & Available & Available & Available & Available \\ \hline
    Low (D $\le$ 0.5m) & Inaccessible & Failed & Failed & Failed & Failed \\ \hline
    Medium (0.5m $<$ D $\le$ 1.2m) & &&&&\\ \hline
    High (D $>$ 1.2m) & &&&& \\ \hline
    \end{tabular}%
  }
  \caption{\label{states}Variables (nodes) and respective states of the BN. }
\end{table}

Next, the CPTs are filled with probability values, starting with the CPT attached to node \textit{Power Supply Generator}. Here, four probability values are required, one for each state of node \textit{Flood Depth at Hospital} (None, Low, Medium, and High), e.g. the probability of the power supply by the hospital failing given a medium flood depth: $P(Power Supply Generator:Failed|Flood Depth at Hospital: Medium)=0.75$. In the case study, we use exemplary failure probability values that increase with rising flood depth (Table \ref{cpt_generator}). The second CPT is attached to node \textit{Power Supply} with its parent nodes \textit{Power Supply Grid} and \textit{Power Supply Generator}. This sub-network expresses an AND gate (see "Methods"), i.e. both parent nodes must fail for the child node to fail as well. Thus, no further information is required for the CPT (see Table \ref{cpt_power}). The third CPT (attached to node \textit{Emergency Care}) shows three parent nodes (\textit{Flood Depth at Hospital}, \textit{Power Supply}, and \textit{Accessibility}). This sub-network represents an OR gate with an additional uncertainty on the impact of node \textit{Flood Depth}. Thus, if either the hospital is inaccessible or the power supply failed, the emergency care service definitely fails. In cases of available power supply and accessibility of the hospital, the impact of the flood depth is uncertain for states \textit{Low} and \textit{Medium}. For this constellation of node states, we again use exemplary failure probability values that increase with increasing flood depth (Table \ref{cpt_hospital}).  

\begin{table}[!ht]
    \centering
    \caption{CPT of node \textit{Power Supply Generator}.}
    \label{cpt_generator}
    \begin{tabular}{ccc}
        \toprule
        &  \multicolumn{2}{c}{\textbf{Power Supply Generator}} \\
        \cmidrule(l){2-3}
        \textbf{Flood Depth at Hospital} & \multicolumn{1}{c}{\small Available} & \multicolumn{1}{c}{\small Failed} \\
        \cmidrule(r){1-1}\cmidrule(l){2-3}
        \small None & 1 & 0 \\
        \small Low & 0.75 & 0.25 \\
        \small Medium & 0.25 & 0.75 \\
        \small High & 0 & 1 \\
        \bottomrule
    \end{tabular}
\end{table}

\begin{table}[!ht]
    \centering
    \caption{CPT of node \textit{Power Supply}.}
    \label{cpt_power}
    \begin{tabular}{cccc}
        \toprule
        &&  \multicolumn{2}{c}{\textbf{Power Supply}} \\
        \cmidrule(l){3-4}
        \textbf{Power Supply Grid}&\textbf{Power Supply Generator} & \multicolumn{1}{c}{\small Available} & \multicolumn{1}{c}{\small Failed} \\
        \cmidrule(r){1-1}\cmidrule(r){2-2}\cmidrule(l){3-4}
        \small Available&\small Available & 1 & 0 \\
        \small Available&\small Failed & 1 & 0 \\
        \small Failed&\small Available & 1 & 0 \\
        \small Failed&\small Failed & 0 & 1 \\
        \bottomrule
    \end{tabular}
\end{table}

\begin{table}[!ht]
    \centering
    \caption{CPT of node \textit{Emergency Care}.}
    \label{cpt_hospital}
    \begin{tabular}{ccccc}
        \toprule
        &&&  \multicolumn{2}{c}{\textbf{Emergency Care}} \\
        \cmidrule(l){4-5}
        \textbf{Flood Depth at Hospital}&\textbf{Power Supply}&\textbf{Accessibility} & \multicolumn{1}{c}{\small Available} & \multicolumn{1}{c}{\small Failed} \\
        \cmidrule(r){1-1}\cmidrule(r){2-2}\cmidrule(r){3-3}\cmidrule(l){4-5}
        None & \small Available &\small Accessible & 1 & 0 \\
        None & \small Available &\small Inaccessible & 0 & 1\\
        None & \small Failed &\small Accessible & 0 & 1 \\
        None & \small Failed &\small Inaccessible & 0 & 1 \\

        Low & \small Available &\small Accessible & 0.75 & 0.25 \\
        Low & \small Available &\small Inaccessible & 0 & 1 \\
        Low & \small Failed &\small Accessible & 0 & 1 \\
        Low & \small Failed &\small Inaccessible & 0 & 1 \\

        Medium & \small Available &\small Accessible & 0.25 & 0.75 \\
        Medium & \small Available &\small Inaccessible & 0 & 1 \\
        Medium & \small Failed &\small Accessible & 0 & 1 \\
        Medium & \small Failed &\small Inaccessible & 0 & 1 \\

        High & \small Available &\small Accessible & 0 & 1 \\
        High & \small Available &\small Inaccessible & 0 & 1 \\
        High & \small Failed &\small Accessible & 0 & 1 \\
        High & \small Failed &\small Inaccessible & 0 & 1 \\
        \bottomrule
    \end{tabular}
\end{table}

\subsection{Spatial Service Models}
Each leaf-node of the hospital-specific service dependency model (dashed nodes in Fig. \ref{bn_study}) is informed by a specific spatial model. The evidence for these nodes can take two forms: regular evidence, i.e. a binary value indicating whether the service is available or unavailable (e.g. node \textit{Accessibility} is in state \textit{Inaccessible}); or soft evidence, i.e. a probability ratio of service availability \citep{Mrad2015} (e.g. the probability ratio of node \textit{Power Supply Grid} available versus unavailable is (0.8,0.2)). Node \textit{Flood Depth at Hospital} is informed by the extreme flood raster layer (Fig. \ref{scenario}) and does not require an additional model. The spatial service models to inform the nodes \textit{Power Supply Grid} and \textit{Accessibility} of each hospital-specific service dependency model are outlined in the following.

\subsubsection{Accessibility Model} \label{accessibility}
\begin{figure}[!ht]
	\centering \includegraphics[width=8cm]{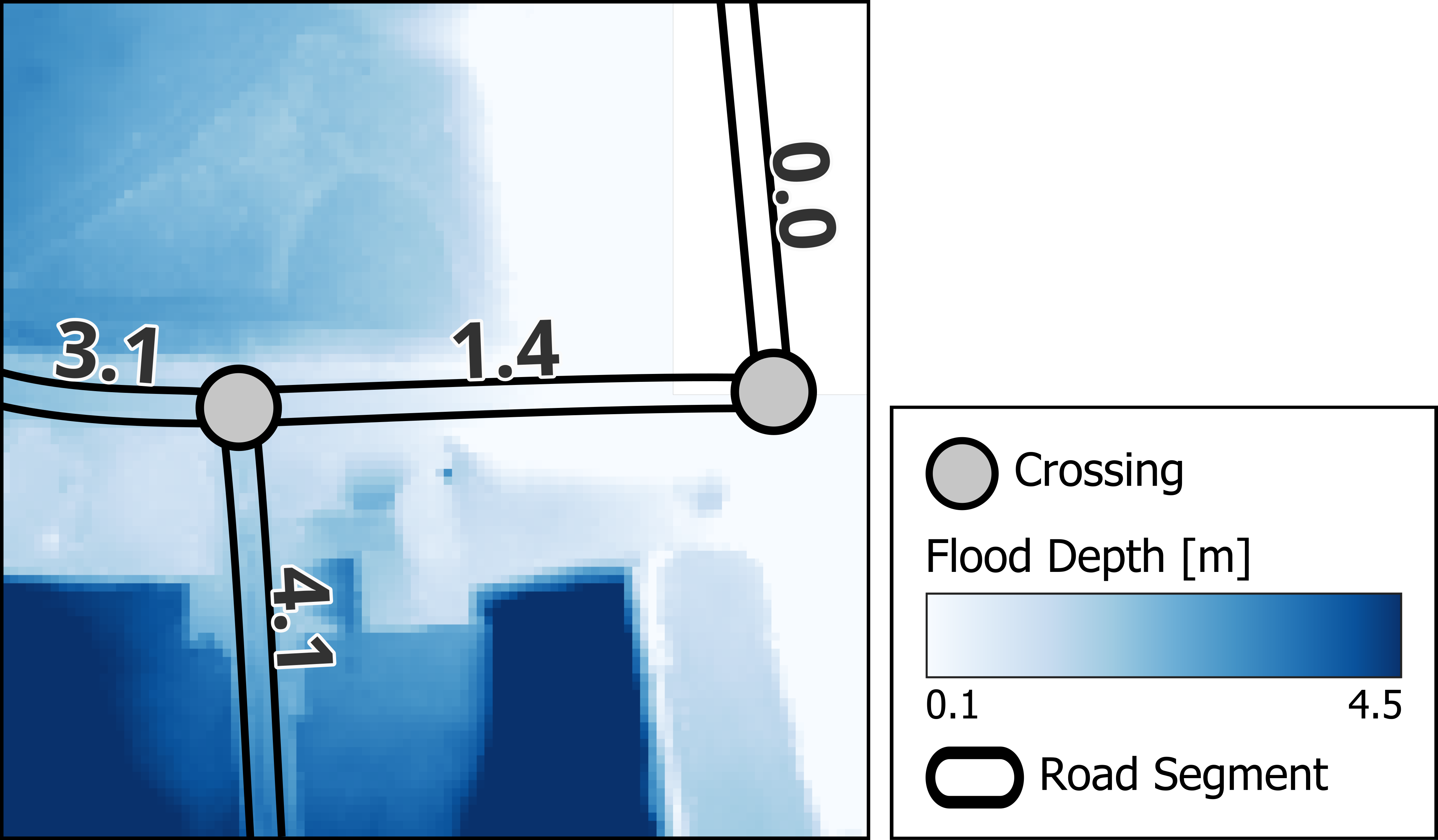}
	\caption{Example area including the reconstructed road network and flood depth layer. Nodes represent crossings and edges represent road segments that include the maximum flood depth at each segment. Additionally, the flood depth raster layer is illustrated that is used to obtain the maximum flood depth at each road segment.} \label{fig:sub1}
\end{figure}

To evaluate hospital accessibility in the study area, the road network is reconstructed using Open Street Map (OSM) data \citep{OpenStreetMap}. The resulting network topology features nodes representing road crossings and edges representing road segments \citep{Li2023} (Fig. \ref{fig:sub1}). Edges are characterized by their length and the maximum flood depth along the section, which is determined from the flood scenario raster layer. In this model, hospital accessibility is represented by two distinct states: \textit{Accessible} and \textit{Inaccessible}. We assume that each road segment in the network can be used (ignoring e.g. one-way streets) and hospitals are accessible if at least one route to the hospital exists where the flood depth for all passed road segments is below 27$\,$cm. This threshold value is based on studies which suggest that a flood depth between 27$\,$cm and 30$\,$cm is the critical point for safely passing through a flooded road segment \citep{Li2023, Gangwal2022}. We argue that these assumptions are sound for emergency scenarios, specifically when considering exclusively emergency response vehicles such as ambulances and fire trucks.

For the routing algorithm, we defined multiple source locations on the east and west side of the Rhine river with sufficient distance from flooded areas. To minimize potential biases introduced by the source location, multiple locations are necessary. For instance, a single source location at the west side of the river might not reveal a feasible path to a hospital, whereas an alternative source location at the east side of the river would demonstrate a viable route. A hospital is considered accessible if a route from at least one source location exists that fulfills the aforementioned criterion. For each source node, the routing algorithm is applied to each hospital in the study area. The outcomes, categorized into the two states (\textit{Accessible} and \textit{Inaccessible}), are then calculated for each hospital to inform the hospital-specific service dependency models.

\subsubsection{Power Supply Grid Model} \label{voronoi}
The power supply of a city depends on the proper functioning of substations, transmission lines, distribution lines, and transformer stations. Substations are crucial hubs that transform high-voltage electricity from the transmission lines to medium voltage levels that are distributed within the city by distribution lines. Transformer stations, which are distributed throughout the city, step down this medium-voltage power to household-friendly voltages. In this case study, we focus for simplicity on the failure of transformer stations caused by the flood. Therefore, we need two models, one for the failure probability of a transformer station given the site-specific flood depth and one that assesses the blackout area if a transformer station fails. 

\begin{figure}[!ht]
	\centering {\includegraphics[width=7cm]{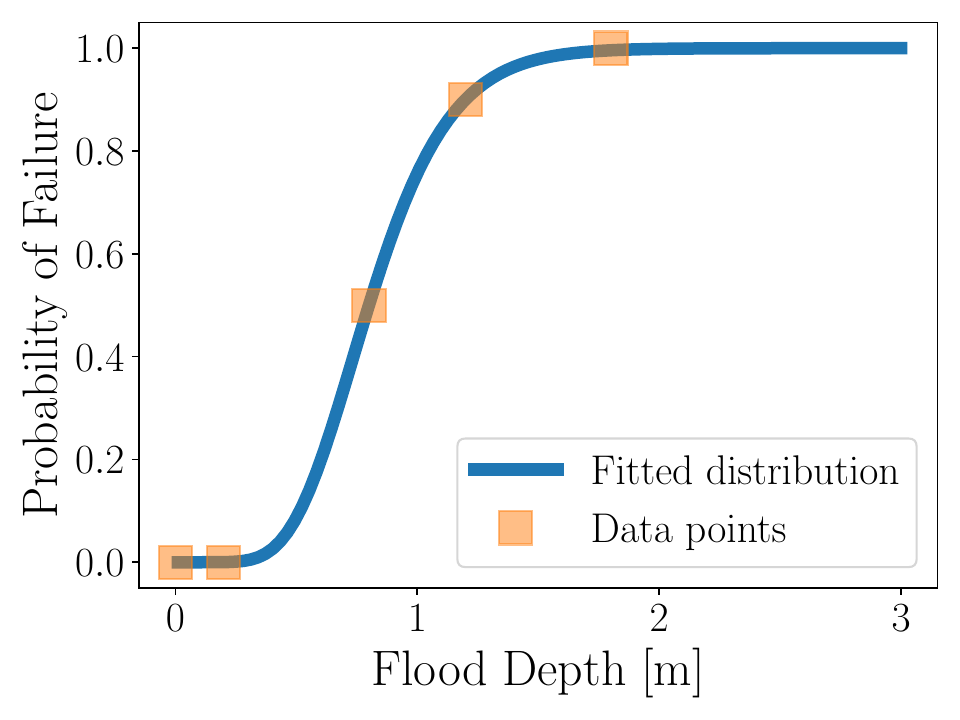}}
	\caption{Fragility function of a transformer station under the stress of a flood. The data points (see Table \ref{data_points}) are used to fit a lognormal cdf using a least-square curve fit algorithm.}
	\label{fragility function} 
\end{figure}

To estimate the failure probability of a transformer station, we use a fragility curve. Fragility curves illustrate the probability of a system or component failing as a function of a given impact measure (IM) of a hazard. Thus, fragility curves show how likely a system is to fail, i.e. become "fragile", under different levels of stress. They can be based on empirical data, simulations, and expert knowledge. If no sufficient empirical data is available, expert knowledge should be included \citep{Mosleh1986,Pita2021}. Here, we use a fragility curve to describe the relation between the flood depth (IM) and the failure probability of a transformer station. \cite{Pita2021} conducted a study on the development of a fragility function for building damages under stress of a flood, i.e. the flood depth for a given period of exposure. Based on empirical data, they showed that the fragility of a building in a flood scenario follows a lognormal cumulative distribution function (lognormal cdf). When empirical data was missing, they used anchor points elicited from experts to fit a lognormal cdf distribution. We follow their approach and use single data points which can be queried by experts (see Table \ref{data_points} for example queries and associated data points) and a non-linear least squares curve fit for the fragility curve (Fig. \ref{fragility function}).

\begin{table}[!ht]
    \centering
    \begin{tabular}{|l|c|c|} 
    \hline
    \textbf{Query} & \textbf{Probability of Failure} & \textbf{Flood Depth}\\
    \hline
    - & 0\% & 0.0m \\ \hline
    No damage up to what depth? & 0\% & 0.2m \\ \hline
    Highest uncertainty of damage at what depth? & 50\% & 0.8m \\ \hline
    Very likely damaged from what depth? & 90\% & 1.2m \\ \hline
    Surely damaged from what depth? & 100\% & 1.8m \\ \hline
    \end{tabular}
    \caption{\label{data_points}Example data points for the fragility function of a transformer station of the case study. Example queries are listed that can be used to elicit a set of probability values and flood depth values serving as data points to fit the fragility curve.}
\end{table}

\begin{figure}[!ht]
	\centering \includegraphics[width=8cm]{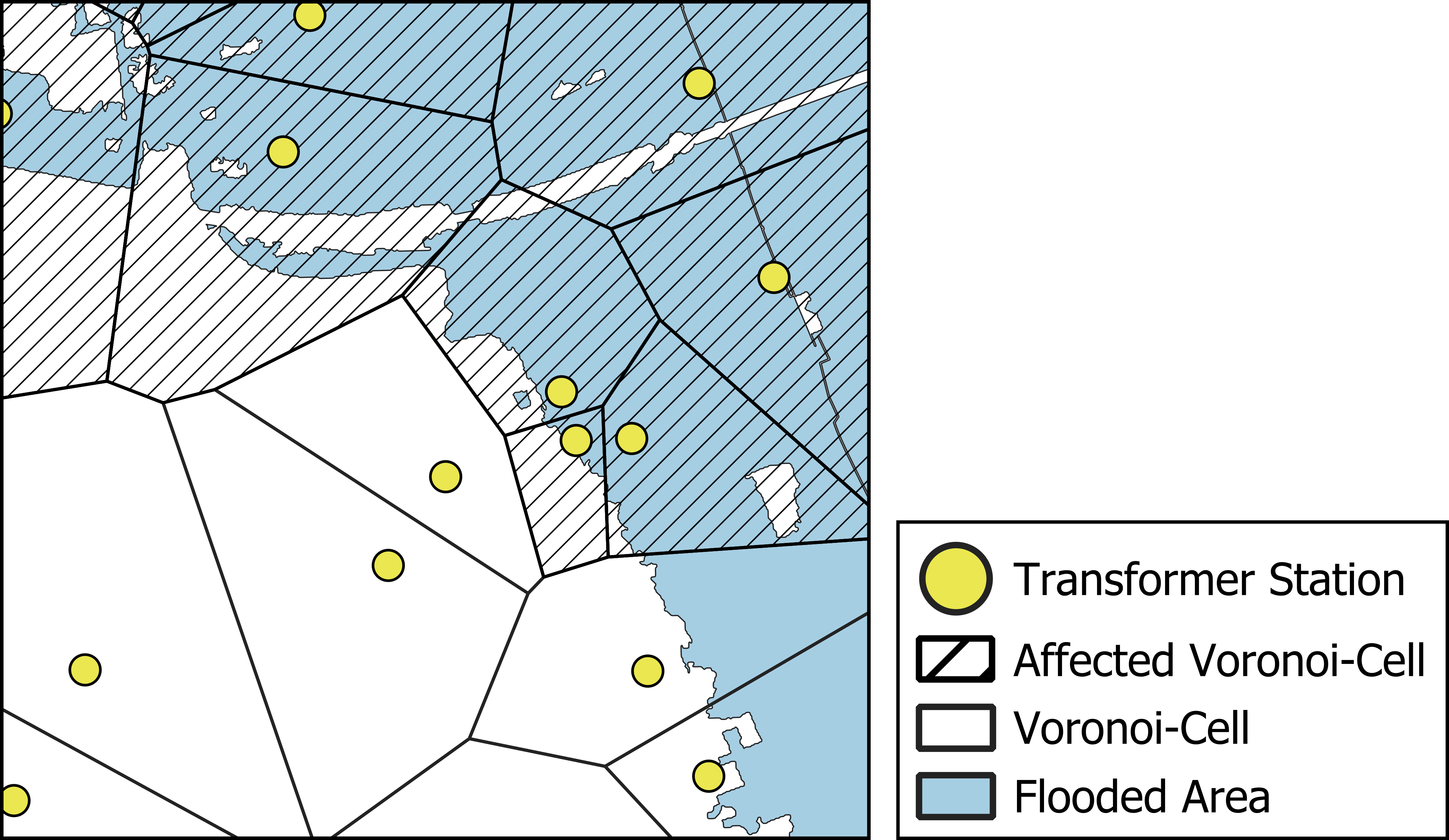}
	\caption{Example area and Voronoi cells affected by the flooding. Transformer stations serve as the source points for the Voronoi diagram. Transformer stations situated within the flooded area are classified as affected, along with their associated Voronoi cells (highlighted with dashed fill).} \label{fig:sub2}
\end{figure}

To identify a potential blackout area for each transformer station, we adapt the Voronoi diagram approach presented by \cite{Held2004}. Voronoi diagrams partition space into regions called cells, each representing the area closest to a source point among a set of given points. In this study, transformer stations serve as the source points, and each cell corresponds to the area nearest its respective transformer station (Fig. \ref{fig:sub2}). 

\subsection{Results and Visualization}
From the total of 27 hospitals we examined, eight hospitals are located within the flooded area, with flood depths ranging from 1$\,$cm to 2.38$\,$m. Six hospitals are deemed inaccessible, either because their access streets are directly impacted by the flooding or because they are disconnected from the road network due to multiple flooded segments at a greater distance. For five hospitals, power supply is uncertain, i.e. the probability of state \textit{Available} for node \textit{Power Supply} is neither zero nor one. One hospital shows a certain power failure ($P(Power Supply:Available)=0$), i.e. both the supply from the grid and the backup power generator fail. Seven hospitals show a complete failure of emergency care service, caused either by inaccessibility, power supply failure, a high level of flooding, or a combination. Two additional hospitals show uncertainty in the states of the emergency care service.    

In the following, the results for two individual hospitals are outlined in detail, illustrating two cases of special interest: accessibility within the flooded area and inaccessibility outside of the flooded area (Fig. \ref{single_hospital}). Subsequently, a dashboard is outlined which shows the results on a map along with summary statistics (Fig. \ref{dashboard}), helping to gain a quick overview over all hospitals in the study area.

\subsubsection{Single Hospitals}
\begin{figure}[!ht]
	\centering \includegraphics[width=17.5cm]{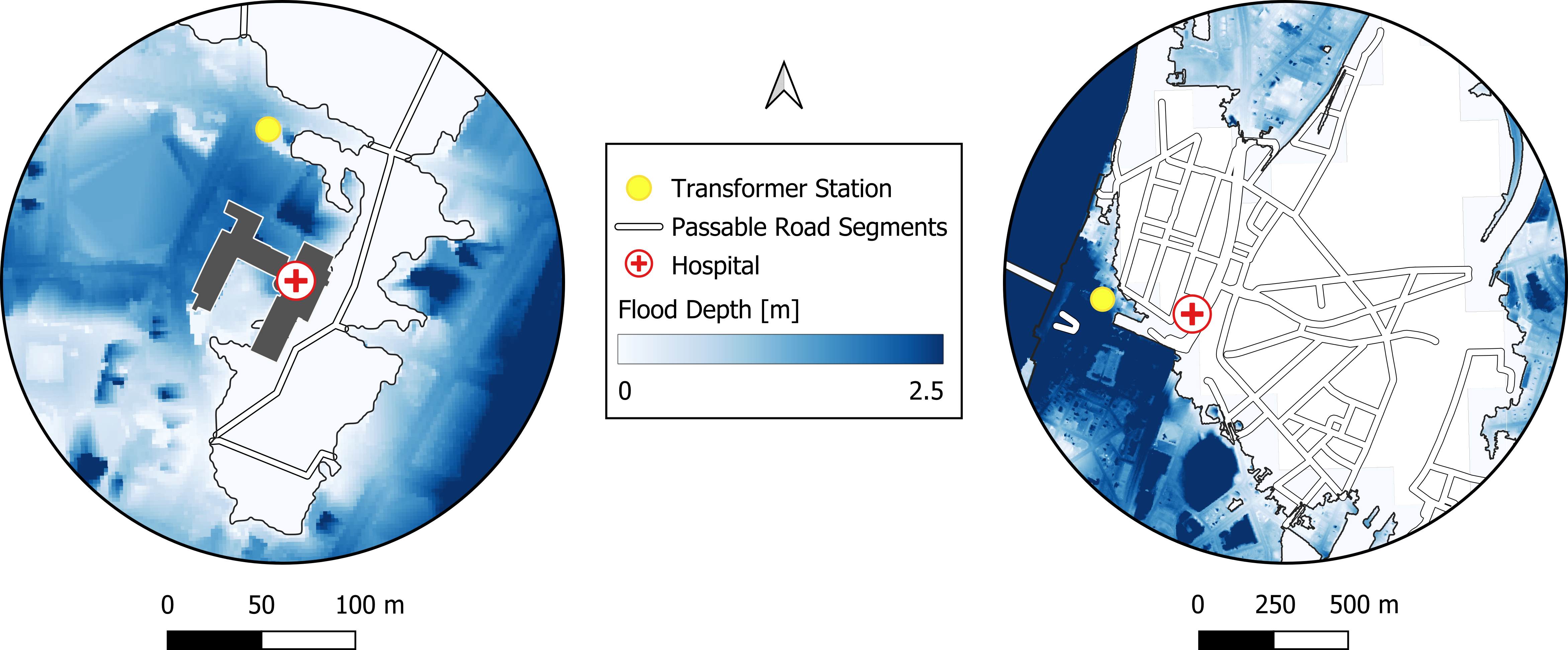}
	\caption{Two example hospitals of the case study. Hospital (1) (see left side) is accessible but within the flooded area (1.12$\,$m). Its respective transformer station is also within the flooded area (0.55$\,$m). Hospital (2) (see right side) is inaccessible but not within the flooded area (0$\,$m), while its respective transformer station is within the flooded area (2.2$\,$m).}
	\label{single_hospital}
\end{figure}

Two selected hospitals illustrate contrasting impacts of the disruption. Hospital (1) (see left side of Fig. \ref{single_hospital}), situated within the flooded area, with a maximum flood depth of 1.12$\,$m, remains accessible despite the flooding. In contrast, hospital (2) (see right side of Fig. \ref{single_hospital}) is not directly affected by the flood but becomes inaccessible due to flooded road segments that disconnect the hospital from the overall road network. The transformer stations for both hospitals are located within the flooded area, with varying water levels: 0.55$\,$m at the station for hospital (1) and 2.2$\,$m for hospital (2). According to the fragility curve (Fig. \ref{fragility function}), these flood depths result in a probability of transformer station failure of 14$\,$\% for hospital (1) and 100$\,$\% for hospital (2). These results are subsequently used to infer the availability of the other services, i.e. nodes \textit{Power Backup Generator}, \textit{Power Supply}, \textit{Emergency Care Service}.

The probability that hospital (1)'s backup power generator is still functional is 25$\,$\%, due to a medium flood depth at the hospital. In combination with the 14$\,$\% of the transformer station failure, this leads to an inferred probability of 90$\,$\% that the hospital possesses reliable power supply. The probability of emergency care service availability is 23$\,$\%. This value is inferred based on the information on nodes \textit{Power Supply}, \textit{Accessibility}, and \textit{Flood Depth at Hospital}. Hospital (2) shows 100$\,$\% service availability of the power supply by the backup generator. This results, due to the OR gate, in a 100$\,$\% service availability of node \textit{Power Supply}. Nevertheless, due to the crucial dependency (modeled by the AND gate) on the accessibility of the hospital for the emergency care services, the resulting probability of emergency care service availability is 0$\,$\%.

\subsubsection{Dashboard}
\begin{figure}[!ht]
	\centering \frame{\includegraphics[width=17.5cm]{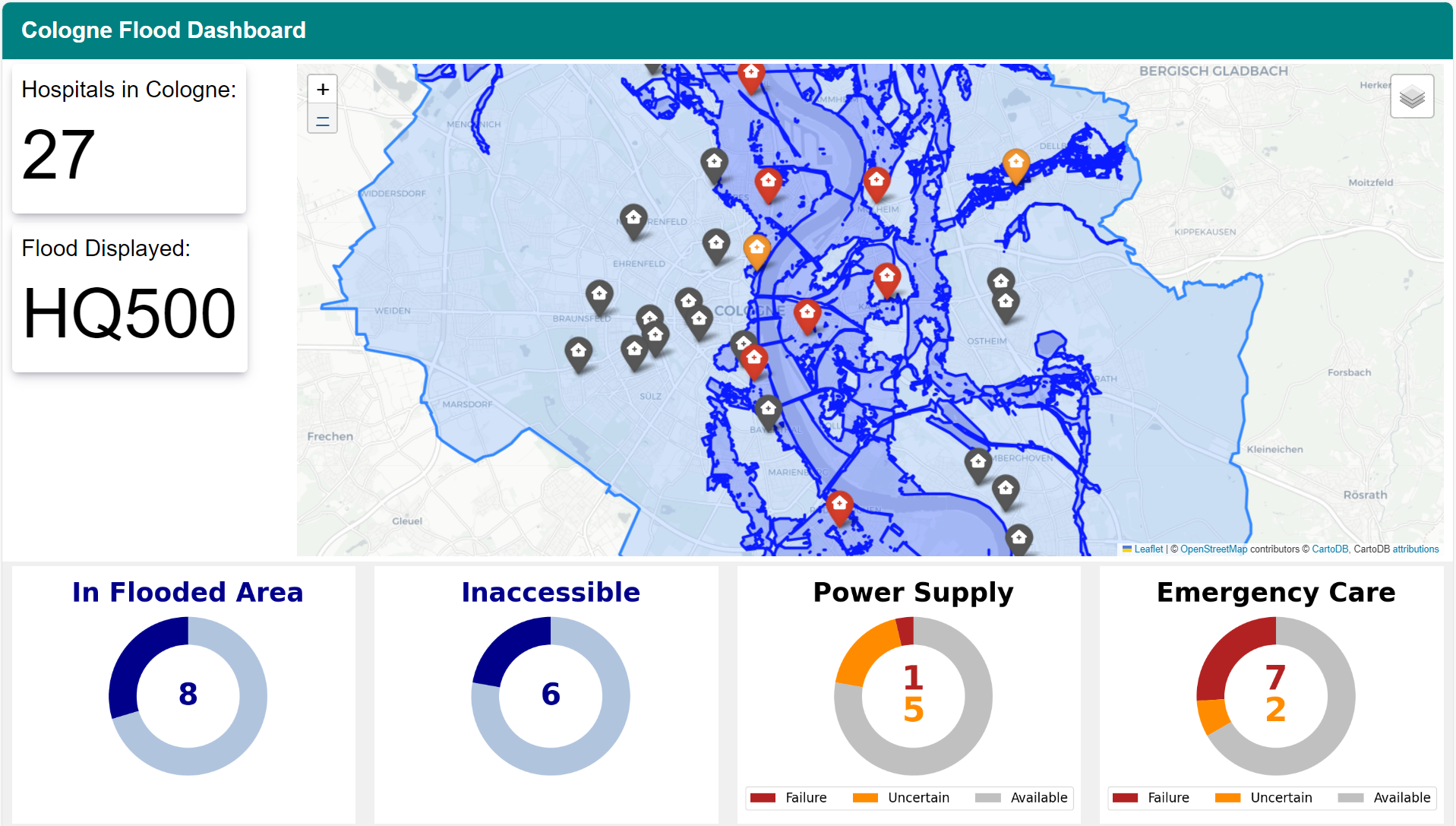}}
	\caption{Dashboard of the case study. The dashboard summarizes the results of the case study for the extreme flood scenario in the city of Cologne. It shows the summarized results of the spatial models (\textit{In Flooded Area} and \textit{Inaccessible} at the bottom left) as well as the results of the hospital-specific service dependency models (\textit{Power Supply} and \textit{Emergency Care} at the bottom right). The map (top right) enables users to explore the impacts on individual hospitals and trace the results of the models.}
	\label{dashboard}
\end{figure}

The dashboard helps to quickly inform decision-makers and facilitates exploratory analysis of the results in an interactive manner. It integrates all insights from the models (spatial models and hospital-specific service dependency models) into a comprehensive and interactive summary (Fig. \ref{dashboard}). At the bottom of the dashboard, spatial model outputs are displayed stating the amount of flooded as well as inaccessible hospitals. In addition, power supply and emergency care service availability is displayed. An interactive map is included, allowing users to trace the results of the hospital-specific service dependency models. In the version of the dashboard shown in this publication, only aggregate data are presented, to avoid conclusions on an individual-hospital level.

\section{Discussion and Conclusion}
In this paper, we introduce \textit{DISruptionMap}, a novel method to spatially assess cascading CI service disruptions in large-scale disaster scenarios. 
Our method consists of three core elements.
First, direct disruptions of CI components, e.g. transformer stations or road segments, are assessed using a GIS-based overlay analysis of CI component locations and a hazard map \citep{Arvidsson2023}. 
The approach allows to consider uncertainties in the direct CI component disruption by using a fragility curve that describes the probabilistic relation between an impact measure (IM) obtained from the hazard map and potential failure states of (CI) components (see Fig. \ref{fragility function}). 
%
Second, the spatial impact of component failures on the corresponding CIs and the area they supply is assessed. 
This is achieved via CI system models which allow a spatially explicit determination of CI service availability in dependence on component failure, e.g. which areas are no longer accessible if certain road segments are flooded (Fig. \ref{fig:sub1}) or which areas are no longer provided with power if certain transformer stations have failed (Fig. \ref{fig:sub2}).
%
Third, indirect (cascading) disruptions are assessed, i.e. the failure of one CI leads to a failure of another dependent CI.
For this, we introduce a Bayesian network-based CI dependency model that is replicated for each target CI (TCI) in the study area -- a procedure similar to the method presented by \cite{Schneider2024}. 
To enable a uniform description of CI dependencies across different sectors \citep{Stergiopoulos2016}, we follow a service-centered approach: a TCI is described by the service it provides for society (a hospital provides emergency care) while other CIs are described by the service they provide for other CIs (a generator provides power for the hospital).
In this way, a spatial assessment of indirect service disruptions on the TCI can be conducted for each TCI entity, e.g. all hospitals in a city.
By combining the GIS-based component failure models and spatial CI models with the BN-based CI service dependency model (GIS informs BN \citep{Johnson2012}), both direct and indirect CI service disruptions can be assessed.

A central motivation for the development of \textit{DISruptionMap} was practicability \citep{Arvidsson2023}.
We aimed for a maximally intuitive work flow based on a minimum amount of additional information, compared to what is required for classic hazard maps.
The resulting approach relies on data available from standard data sources -- i) static hazard maps are readily available for different hazard scenarios \citep{Geiss2022,Fekete2019}, ii) CI locations can be collected from OSM, web services provided by governmental registry offices, or they can be obtained directly from local CI providers, and iii) expert knowledge can easily be integrated into the analyses, e.g. for setting up the service dependencies within the BN-based model. 
To facilitate the integration of expert data, we use a fault tree for creating the CPTs of the service dependency model \citep{Bobbio2001,Rahimdel2024}. 
The fault tree-based approach reduces the amount of input information required for the CPTs, by using logical operators (i.e. gates) to describe the conditions for service availability. This significantly reduces the workload for the consulted experts \citep{Druzdzel2000}.

In its current form, \textit{DISruptionMap} relies on four main simplifying assumptions for incorporating service dependencies in the analysis of disaster risks. If needed, this basic framework can be easily extended, depending on the requirements of the chosen use case.
%
For instance, instead of using binary variables, a higher granularity of service availability states could be implemented (e.g. \cite{Grafenauer2018} use a five-point scale between service availability and failure). However, this would require much more data to set up the BN and a more detailed assessment of direct hazard impacts, e.g. multiple fragility curves, one for each failure state. 
%
Another potential extension concerns the uncertainty in the effect of the failure of one infrastructure on the service provision of another. 
For instance, a power outage certainly causes traffic impairments \citep{Rehak2018}, but to what extent the traffic system is still functional without power depends on many other factors such as the amount of traffic that has to be regulated by traffic lights. 
In our \textit{DISruptionMap} method, the implementation of uncertain service dependencies is rather straightforward due to the probabilistic nature of BNs: the deterministic (AND and OR) gates in the fault tree procedure could be replaced by probabilistic or noisy gates \citep{Bobbio2001}.
However, the practicality and effectiveness of including such gates should first be tested with potential end users as this implies a higher workload for the consulted experts.

A potential extension which is less straightforward to implement is the consideration of the time dimension. In reality, both the evolution of a hazard scenario and the characteristics of the induced cascading (infrastructure) service failures show temporal dynamics \citep{Arvidsson2023, Witte2021}. For example, the flood depth at critical road segments may vary over time and fuel-based emergency power generators operate for limited time spans, depending on fuel reserves. First of all, including the time dimension would require to explicitly describe how infrastructure dependencies vary over time, which is a time-consuming and data-intensive task (see, e.g. \cite{Stergiopoulos2016} or \cite{DiGiorgio2012} for similar approaches, which do not explicitly consider the spatial dimension). Furthermore, this extension would require a dynamic map for specific hazard scenarios, e.g. a dynamic simulation of flood propagation. Such simulations are costly in their development, not likely to be available for various hazards, and would thus drastically reduce the applicability of the method. 
%
Lastly, it would certainly be desirable to extend our method to include CI \textit{inter}dependencies. While a CI dependency refers to a one-directional relation of the state of one CI on the state of another CI, an interdependency describes a bi-directional connection (i.e. a feedback loop) between two CIs \citep{rinaldi2001}. 
However, in a BN, such a feedback loop cannot be directly represented since a BN is by definition acyclic (i.e. it does not contain loops). 
A work-around to include interdependencies in a BN-based method could be to use the output of the BN-based service dependency models again as an input for the spatial service models -- a complex procedure that would require much more data, which is difficult to elicit from experts.  

To conclude, the \textit{DISruptionMap} method balances the level of detail in modeling complex CI service disruptions during large-scale disaster scenarios and the amount of data required to set up the corresponding models.  
Building \textit{DISruptionMap}, we made sure that the method mainly relies on data that is easy to acquire and that its structure enables a straightforward elicitation and integration of missing data.
Also, \textit{DISruptionMap} is highly versatile: it can easily be applied to various hazard scenarios, sets of services, and study areas.
Therefore, the method is suitable for local disaster management authorities that already use hazard maps, have access to the standard data sources, and possess the professional expertise to set up the models. 
For these authorities, \textit{DISruptionMap} provides a low-threshold upgrade to existing hazard maps that allows to incorporate cascading effects in the analysis of and the preparation for disaster risks. 

\section{Methods} \label{methods}
\subsection{Infrastructure Service Dependency Model}
The development of the BN-based CI service dependency model is divided into two steps that ensure a systematic model construction for a specific disaster scenario and region: (i) \textit{Graph Development}, starting from the target critical infrastructure (TCI) and the impact measure for the hazard under consideration and (ii) \textit{Quantification of Service Dependencies}, based on expert knowledge. 

\subsubsection{Graph Development} \label{graph_structure}
\begin{figure}[!ht]
	\centering \includegraphics[width=9cm]{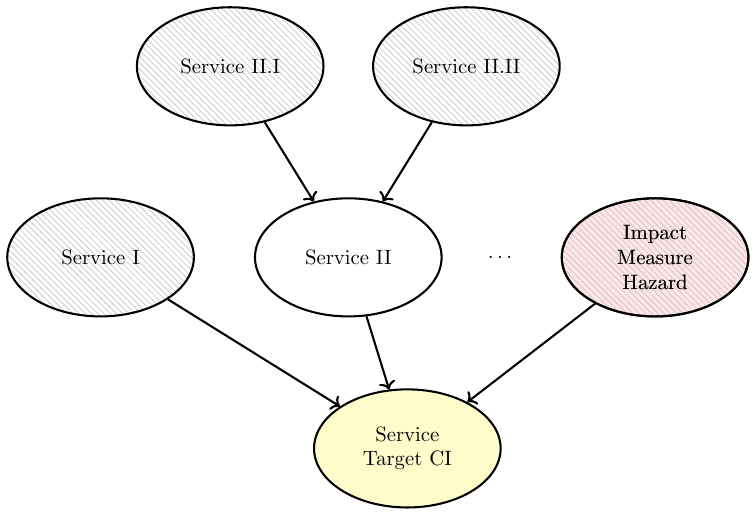}
	\caption{Generic graphical structure of the method. The graph centers around a target CI (TCI) service variable (highlighted with yellow fill) which serves as the starting point of an iterative approach to identify service dependencies. The node representing the impact measure of the hazard (highlighted with red fill) is introduced as a parent node of the target node. Additionally, the graph features nodes that are solely informed by other nodes (white fill) and leaf nodes that are informed by the spatial service models (dashed fill).}
	\label{general_bn}
\end{figure}

The proposed method centers around a TCI, serving as the starting point for an iterative approach which seeks to identify dependencies of the TCI on other services. Starting from the TCI, we ask 'What services must be present for the TCI to function?'. Each identified service is represented by a node that is linked to the TCI (Fig. \ref{general_bn}). Each service can be further detailed, i.e. again serving as a starting point for the proposed inquiry. This iterative process should be repeated until all leaf service nodes (i.e. nodes with no incoming edges, represented by the dashed nodes in Fig. \ref{general_bn}) show a predominant direct disruption by the hazard.
The leaf nodes themselves are later informed by the spatial service models. In addition, one variable describing the impact measure (IM) is added as a child node to the node \textit{Service TCI} (Fig. \ref{general_bn}) to account for the disruption of the TCI service due to the hazard. In this way, the service dependency model accounts for direct disruptions (see leaf nodes in Fig. \ref{general_bn}) and indirect disruptions (see \textit{Service II} in Fig. \ref{general_bn}) on the TCI service.

\subsubsection{Quantification of Service Dependencies} \label{method_details}
The dependencies among individual services are defined via conditional probability tables (CPTs) within the BN. For each dependent node (see node \textit{Service II} and \textit{Service Target CI} in Fig. \ref{general_bn}), a CPT is required that includes the probabilities for all combinations of parent and child node states. To estimate the respective probabilities, there are generally two options: historical data and expert knowledge. The availability of historical data for large-scale hazard scenarios that caused serious cascading events is often limited. Therefore, in this work, we use expert knowledge to build an accurate and reliable BN despite the absence of sufficient empirical evidence. We propose the utilization of a fault tree approach as a straightforward and effective means to describe dependencies that keeps the required information for setting up the CPTs at a minimum. Fault-trees can be translated into BNs \citep{Bobbio2001,Rahimdel2024}, and are thus easily integrated into the proposed BN-based method. 

\begin{figure}[!ht]
	\centering \includegraphics[width=16cm]{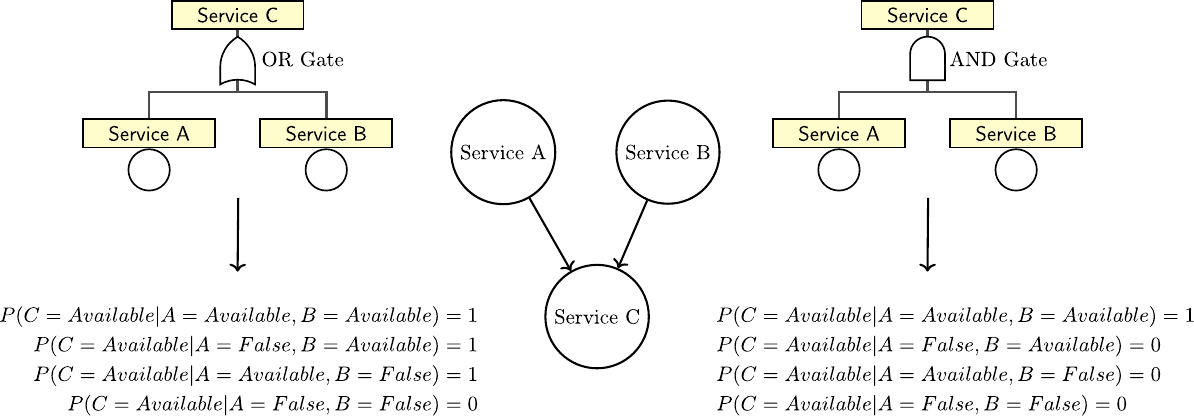}
	\caption{Example of an OR Gate (left side) and an AND Gate (right side) in Fault Tree and BN (middle) representation (adapted from \cite{Bobbio2001}). Both fault trees and BN show the same set of services (nodes) and the same structure (edges). Below the Gate representations, the respective translation of the Gate into the probability values for the dependent node \textit{Service C} is illustrated.}
	\label{fault_tree}
\end{figure}

Fault trees require a graphical model that shows all possible paths leading to a specific failure. The directed acyclic graph (DAG) of the BN mirrors this fault tree structure (Fig. \ref{fault_tree}). In fault tree analysis, AND and OR gates are used to combine the effect of (service) failures into more complex configurations, enabling the identification and quantification of potential failure paths in a system. AND gates require that all input services (e.g. service A and B in Fig. \ref{fault_tree}) are functional in order to provide the output service (service C). They are represented by a $\cap$ symbol. OR gates, on the other hand, require that at least one input service is functional for the output service to be available. They are represented by a $\cup$ symbol. Given a BN configuration of only one child node with its parent nodes (e.g. see Fig. \ref{fault_tree}), an expert only needs to provide one piece of information: whether this combination is best represented using an AND or an OR gate. Given this decision, the respective CPT attached to the child node can be automatically populated (constituting eight probability values in the example of Fig. \ref{fault_tree}). 

\subsection{Spatial Models} \label{connection}
The spatial models are highly dependent on the services being modeled. Nevertheless, they share certain requirements that need to be fulfilled to provide the output required for the TCI-specific service dependency models. The BN-based CI service dependency model is replicated and assigned to each TCI in the study area (see \textit{ERIMap} method \citep{Schneider2024}), e.g. all fire stations in a specific district or all hospitals in a city. These TCI-specific BNs are then informed by the spatial service models. These spatial service models must meet three key requirements: (i) they must provide detailed, spatially explicit information about the status of each infrastructure service. This includes, for instance, which locations are inaccessible due to road network disruptions. (ii) The information provided must match the requirements of the leaf nodes in the BN. For example, if a node describes power supply by the power grid, the spatial service model must explicitly indicate the affected areas. (iii) The output of these models must be either unambiguous (the service states \textit{Available} or \textit{Failed}; this is processed as hard evidence in the BN) or constitute a probability ratio (e.g. $P(Available)=0.8$ and $P(Failed)=0.2$; this is treated as soft evidence, see \cite{Schneider2024}). 

\bibliographystyle{chicago}

\end{document}